\RequirePackage{fix-cm}

\documentclass[twocolumn]{svjour3}          
\smartqed  
\usepackage{graphicx}
\usepackage{epstopdf}
\usepackage{enumerate}
\usepackage{amsmath}
\usepackage{booktabs}
\usepackage{color}
\usepackage{natbib}
\usepackage{amsfonts}
\usepackage[normalem]{ulem}

 \journalname{Exp Fluids}

\begin{document}

\title{Impaction of spray droplets on leaves: influence of formulation and leaf character on shatter, bounce and adhesion}

\titlerunning{Impaction of spray droplets on leaves}        

\author{Gary J. Dorr \and
        Shuangshuang Wang \and
        Lisa C. Mayo \and
        Scott W. McCue \and
        W. Alison Forster \and
        Jim Hanan \and
        Xiongkui He
}

\institute{G.J. Dorr (joint first author), J. Hanan
\at The University of Queensland, Queensland Alliance for Agriculture and Food Innovation, Brisbane, Australia
\and
S. Wang (joint first author), X. He
\at Centre for Chemicals Application Technology, China Agricultural University, Beijing 100193, China \\ \email{xiongkui@cau.edu.cn}
\and
L.C. Mayo, S.W. McCue
\at Mathematical Sciences, Queensland University of Technology, Brisbane, Australia \\ \email{scott.mccue@qut.edu.au}
\and
W.A. Forster
\at Plant Protection Chemistry NZ Ltd, Rotorua, New Zealand
}

\date{\today}

\maketitle


\begin{abstract}
This paper combines experimental data with simple mathematical models to investigate the influence of spray formulation type and leaf character (wettability) on shatter, bounce and adhesion of droplets impacting with cotton, rice and wheat leaves.  Impaction criteria that allow for different angles of the leaf surface and the droplet impact trajectory are presented; their predictions are based on whether combinations of droplet size and velocity lie above or below bounce and shatter boundaries.  In the experimental component, real leaves are used, with all their inherent natural variability.  Further, commercial agricultural spray nozzles are employed, resulting in a range of droplet characteristics.  Given this natural variability, there is broad agreement between the data and predictions.  As predicted, the shatter of droplets was found to increase as droplet size and velocity increased, and the surface became harder to wet. Bouncing of droplets occurred most frequently on hard to wet surfaces with high surface tension mixtures.  On the other hand, a number of small droplets with low impact velocity were observed to bounce when predicted to lie well within the adhering regime.  We believe this discrepancy between the predictions and experimental data could be due to air layer effects that were not taken into account in the current bounce equations. Other discrepancies between experiment and theory are thought to be due to the current assumption of a dry impact surface, whereas, in practice, the leaf surfaces became increasingly covered with fluid throughout the spray test runs.
\end{abstract}


\section{Introduction}
\label{sec:intro}

Pesticides are used for the management of weeds, insects or pathogens in most agricultural cropping systems worldwide.  They are generally applied as a spray with the aim of covering all or part of the target (pathogens, insects, leaves or other plant parts).  To be effective at controlling the pest, the droplets must be retained on the target surface and not lost as spray drift, or deposited on the ground and other non-target sites within the sprayed area.  Within this context, we present here experimental data on droplet impaction events using real leaves, commercial spray nozzles, and a variety of spray adjuvants.

While our motivation is to study droplet impaction in the context of agrichemical spraying, another closely related application of this type of work is the spreading of pathogens from leaf to leaf by rain droplets after impaction.  Splash from rainfall or overhead irrigation is important in the dispersal of particles from natural surfaces, since secondary drops produced when rain droplets impact a surface can pick up pathogens and deliver them to another plant. Rain-splash is probably the second most important natural agent for dispersal of spores after wind \citep{Huber1997,Fitt1989};  splashing provides an effective short-range dispersal, particularly when the leaf canopy is dense, whereas wind is the mechanism behind long-range dispersal \citep{Calonnec2013}.  In contrast to pesticide application, droplet sizes from rainfall are much larger in size. Typically the diameters of rain drops are greater than 1mm, whereas for pesticide application median diameters are typically less than 0.5mm \citep{Dorr2013}.  Further, in a rainfall event, leaves are typically wet and hence are more prone to shatter when compared to field application of pesticides, where most of the droplets impact upon a dry area of the leaf. Nevertheless, the shatter events which occur during either pesticide application or via rainfall are largely determined by a competition between inertial and surface tension forces acting on the droplet at impact \citep{SaintJean2006}.

There are three main possible outcomes when a droplet impacts the target: it can adhere, bounce off the surface, or shatter into a number of smaller droplets \citep{Mercer2010}. Each of these three outcomes can take different forms. If a droplet adheres it can return to the shape of the original droplet and rest on the surface, or it can spread over an area much larger than the initial droplet diameter \citep{Xu2011}. If a droplet bounces it can fully rebound or part of the droplet can remain at the point of impact (partial rebound; Yarin, 2006). Shatter can take many different forms, for example prompt splash, corona splash and receding breakup \citep{Yarin2006,Marengo2011,Moreira2010}. Additional impaction behaviours can involve droplets sliding or rolling along the surface \citep{Mayo2015,Veremieiev2014}, or the formation of rivulets, wherein the droplet slides downslope while spreading but recoil does not occur \citep{Antonini2014,Dong2013,Dong2014}. However, for the purpose of this paper we will use the three outcomes as outlined in \citet{Mercer2010}.

When droplets bounce or shatter on impact with a target, the retention of the spray at the position of impact is negligible compared to adhesion on impact.  However, these bounce or shatter daughter droplets can redistribute to other parts of the target or nearby targets \citep{Dorr2013a}.  Alternatively, droplets may bounce or shatter from non-target surfaces onto the target. Thus, each impact event may positively contribute to retention, and it is important to understand when and how each takes place.

There are several key stages common to any single droplet impaction \citep{Attane2007,Mao1997,Mercer2010}. First, the drop impinges on the surface and inertial forces are redirected from the impact trajectory to directions at a tangent to the surface. This initiates a spreading stage where the drop's surface area increases, but kinetic energy is dissipated due to viscous forces within the fluid and at the advancing contact line. If the drop has not shattered in response to the large inertial forces, spread continues until kinetic energy is depleted entirely. In some cases this may be the end of the droplet's journey, with adhesion being the end result. In other cases, particularly when the surface is difficult to wet, the increased surface energy due to spread can cause recoil. If the drop recoils with enough energy, it will rebound from the surface (partially or completely; Rioboo et al. 2001). Alternatively, the drop will undergo several oscillations of spread and recoil but ultimately settle down to an equilibrium position and adhere to the surface.

A common approach to modelling spread and recoil is to consider the changes in kinetic and surface energy as the drop oscillates on the surface. Several authors have developed ordinary differential equation (ODE) models based on energy equations, which describe how the radius of a drop varies during spread and recoil \citep{Attane2007,Kim2001}. Most notably, Attane et al. presented a semi-empirical model based on an assumed rimmed-disk geometry which closely predicted the maximum spread factor for a range of fluid and impact parameters. This model was then extended by \citet{Mercer2010} to incorporate the effect of differing contact angles for advancing and receding motions, and also to include a predictor for bounce.

Such ODE models are to be contrasted with the purely algebraic energy balance model by \citet{Mao1997}, which not only closely predicts maximum spread factor (when compared to various data from the literature; Mao et al. 1997), but also provides a bounce criterion, without the need to solve a nonlinear ODE for each impaction. The reduced computational time of an algebraic bounce model is highly advantageous for large scale applications where thousands of drops may be present at the same time, such as in multi-component simulations of agricultural spray scenarios \citep{Dorr2013a}. While there are models in the literature which can accurately simulate the entire impaction process in three dimensions \citep{Malgarinos2014}, they are very computationally intensive even for a single droplet. It is for this reason that the present work focuses on purely algebraic energy balance models. We employ Mao et al.'s model to describe and predict droplet bounce, with simple modifications to allow for the calculation of rebound velocity, and to apply the model to the case of an oblique impaction.

The mechanisms behind droplet shatter are not yet completely understood \citep{Rein2008}. As a result, much of the literature consists of empirical models for the prediction of shatter based on one theoretical foundation in hydrodynamics: droplet shatter occurs when the inertial forces of impact overcome the capillary effects of the fluid \citep{Moreira2010}. Empirical shatter criteria are typically simple equations involving dimensionless quantities such as the Weber, Reynolds and Ohnesorge numbers, which indicate whether shatter occurs or not \citep{Mundo1995,VanderWal2006,Rein2008}. In addition to the hydrodynamic argument on which these criteria hinge, there is an ongoing interest in the literature in ideas such as the role that air entrapment \citep{Rein2008,Mandre2012,Kolinski2014} and atmospheric pressure \citep{Xu2005,Riboux2014} play in the onset of droplet shatter. Here we simply use the Mundo et al. criterion to predict shatter for normal and oblique impactions \citep{Mundo1995}.

There are few studies which model the \emph{post}-shatter behaviour deterministically (number, velocity and trajectory of secondary drops), although it can be achieved to some extent through energy balance models very similar to those used for spread and bounce \citep{Yoon2006}. Alternatively, empirically-based statistical models may be used, such as those by \cite{SaintJean2004,Gigot2014} for the rain-splash dispersal of pathogens. We do not consider post-impaction events in this study, as these processes are difficult to observe and measure accurately within the current experimental framework.

High speed video images have been used by researchers to study the impaction process, although these have typically been on uniform, well defined surfaces such as glass \citep{Roux2004}, parafilm \citep{Wang2009} or PTFE coated microscope slide \citep{Massinon2012}. \citet{Reichard1986}, \citet{Reichard1998}, \citet{Fox1992}, and \citet{Wirth1991} used high-speed cameras to record droplet impaction on leaves.  The speed and image resolution of these cameras, however, could not capture the dynamic impaction process, and the properties of daughter droplets could not be determined because many droplets overlapped within the camera target focus area. More recently, \citet{Dong2013} and \citet{Dong2014} developed a system that used two high-speed digital cameras and 3D software to study the impact of a range of droplet sizes and impact velocities on \emph{Hydrangea quercifolia}, \emph{Dracaena}, and \emph{Euphorbia pulcherrima}, and was used to qualitatively describe the impaction process and measure the spread of droplets on impact.

In other recent research, \citet{Massinon2012} and \citet{Zwertvaegher2014} studied droplets of both water and surfactant mixtures impacting on horizontal, synthetic, hydrophobic polytetrafluoroethylene, Teflon$^{\circledR}$ (PTFE) coated slides. Drops were delivered to the surface with a moving agricultural nozzle, and impaction details (droplet size and velocity) and outcomes (adhesion, bounce and shatter) were extracted from high speed video images. Their results show a zone of transition between each impaction outcome and not a clear boundary between them. A `Weber number of transition' was used to delineate the boundary between adhesion and bounce, and bounce and shatter. These transition values were obtained from the intersection between the Weber number probability density distributions of their data. In some cases (when surfactant concentration was sufficiently increased), bounce outcomes were not observed at all, and so only one Weber number of transition was required, describing a boundary between adhesion and shatter zones \citep{Massinon2012}.

Likewise, \citet{Massinon2014} obtained similar results for water and surfactant mixtures on PTFE and blackgrass leaves of various orientations. Rather than using Weber numbers of transition to delineate adhesion-bounce and bounce-shatter zones, they reported impact outcome probability as a function of eleven energy classes, where each class boundary corresponds to a constant Weber number. As with the Weber numbers of transition \citep{Massinon2012,Zwertvaegher2014}, \citet{Massinon2014}'s impact outcome probabilities in each energy class were experimentally-determined. Spraying distilled water was found to result in less adhesion, more bounce, less shatter in Cassie-Baxter regime and more shatter in Wenzel regime than when a spray mixture of 0.1\% Break-Thru S240$^{\circledR}$ was applied. For impaction on blackgrass leaves, \citet{Massinon2014} found that the leaf roughness pattern clearly affects drop impact behavior compared with an artificial surface. From these results it seems reasonable to expect that impaction outcomes may be more variable on natural leaf surfaces due to their inherent variability \citep{Nairn2014,Forster2005,Taylor2011}.

Experimental techniques such as those outlined by \citet{Massinon2012}, \citet{Zwertvaegher2014}, \citet{Massinon2014} and \citet{Dong2014} demonstrate a difference in impaction outcomes between different surfaces and spray mixtures. However, it is desirable for future modelling of whole plant spraying, and also pathogen spreading scenarios, to be able to predict these effects without being totally reliant on experimental data for each possible application scenario.

In this paper we report on experimental data taken from high speed video images of droplets impacting on cotton, wheat and rice leaves.  In an attempt to more closely mimic conditions in the field, we use three different commercial agrichemical spray nozzles and four different formulation types (one of which is simply water, while the other three include commonly used adjuvants).  The experimental details are provided in Section~\ref{sec:video_analysis}.  For both horizontal leaves and leaves held at $45^\circ$, the outcome (adhere, bounce or shatter) for each droplet impaction was recorded, as well as the droplet size and speed just before impaction.  The mathematical approaches outlined above (and described in Section~\ref{sec:models}) are tested by plotting bounce and shatter boundaries as a function of droplet size and speed, and comparing with the experimental results (Section~\ref{sec:results}).  Section~\ref{sec:discussion} includes a discussion of our work.  We find broad agreement between the data and predictions, but also highlight cases in which observations are not backed up by the models as they stand.  This research provides confidence that simple and computationally efficient mathematical models can be used to investigate retention of pesticides on whole plant from typical pesticide applications \citep{Dorr2008,Dorr2013a}, but also points to aspects of the modelling which needs further refinement.


\section{High speed video analysis}
\label{sec:video_analysis}


\subsection{Experimental set up}
\label{sec:exp_setup}

\begin{figure}
\includegraphics[width=\columnwidth]{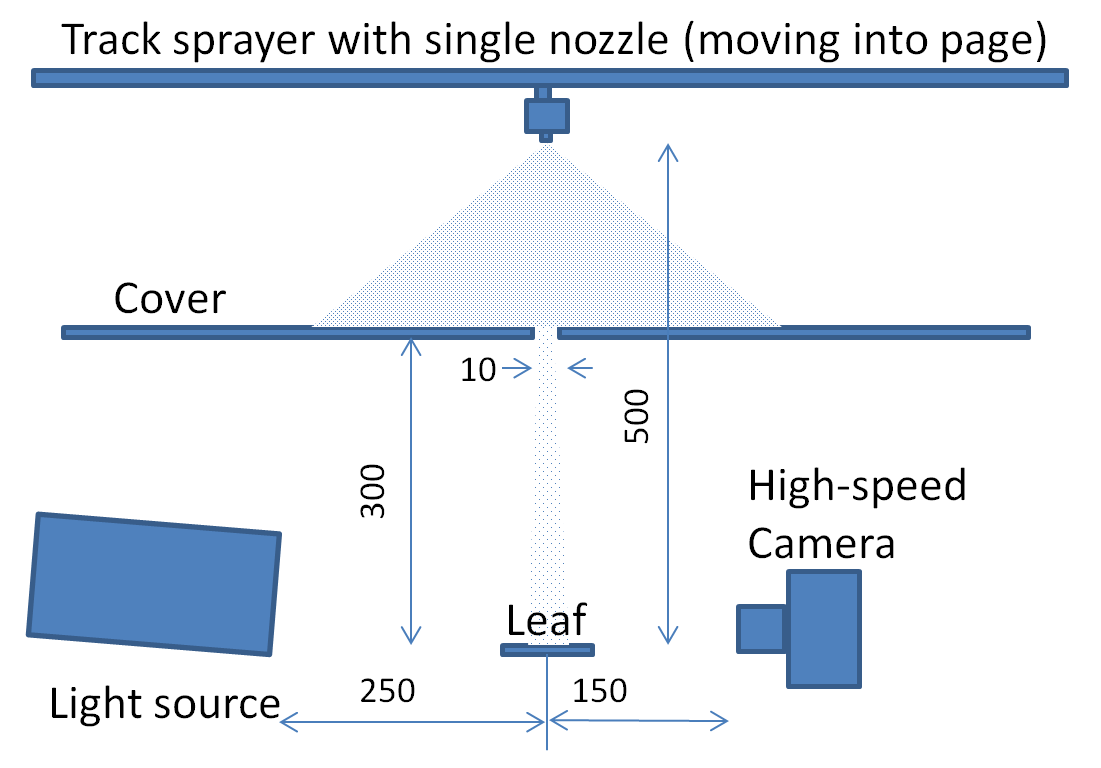}
\caption{Schematic layout of high speed video experiments}
\label{fig:exp_layout}
\end{figure}

Droplets with a range of diameters and velocities were recorded impacting on three plant species with a high speed video system at a frame rate of 5000 f/s and shutter speed of 1/20000 s.  The system consisted of a Lightning RDT $^{\text{Plus}}$ high-speed camera (DRS Technologies, Inc., USA), an HL-250 high-intensity discharge light with maximum power of $1,250$W (Shanghai Anma Industry Co., Ltd,  China) and integrated with MiDAS $2.0$ software modules (Xcitex, Inc). The experimental layout is shown in Fig~\ref{fig:exp_layout}.

During each test run, a target leaf was fixed on a sample board and placed on an aluminum alloy frame that enabled the angle of the leaf to be altered. The camera focused on the leaf surface to capture the impaction process. To protect the camera and light from spray and to limit the number of out of focus droplets, a plastic board with a slot $0.01$ m wide and $0.1$ m long was used as a cover. The slot was positioned over the leaf to allow droplets to go through and impact on the target at the camera focal plane.

By using a slot to limit the number of droplets, together with a low application rate that is typical of field applications, we were able to ensure that the majority of the impacting droplets hit a dry area on the leaf.  To verify that most of the droplets impact an area of the leaf where there was no previous impact, we sprayed water under the same operating conditions on water sensitive cards ($75\times 25$ mm) that change color from yellow to blue when a droplet impacts on their surface.  A figure demonstrating these results is included in the supplementary material.

After recording the spray, the movement of droplets that were in good focus prior to impact was manually tracked using the MiDAS software and the impact velocity of the droplet was calculated. It was then noted from the recorded images if that droplet adhered, bounced or shattered on impact. To measure the droplet size, a public domain image processing program (ImageJ 1.48c, developed at the National Institutes of Health) was used. Towards the end of the frame sequence, droplets that were suspected to have hit a wet area (due to a visible change in behavior of the impacting droplet) were excluded from the analysis.


\subsection{Spray methods and formulations}
\label{sec:spray_methods}

The distance from the nozzle tip to the target was $0.5$ m. The spray pressure was $0.3$ MPa, supplied by a TYW-2 air compressor (Suzhou Tongyi Electromechanical Co., Ltd., China). The droplets were produced by three nozzle types (IDK120-02, ST110-02 and TR80-02 nozzles, Lechler GmbH, Germany) whose respective Volume Median Diameters (VMD) with water were 273 $\mu$m, 174 $\mu$m and 171 $\mu$m, respectively, measured by a SympatecHelios laser-diffraction particle-size analyser (Spraytec laser-diffraction particle-size analyser, Malvern Instruments Ltd., UK) at the stated spray distance and pressure \citep{Wang2014}. The nozzle was mounted on a track sprayer, operated at a constant speed of $1.0$ m/s over the test zone.

The following solutions were tested:
\begin{itemize}
\item Water
\item $0.05\%$ NF100 (Spray adjuvant, Numen (Beijing) International Biotech Co., Ltd., China) $+$ water
\item $0.1\%$ Tween $20$ $+$ water
\item $0.05\%$ Silwet 408 (Spray adjuvant, Numen (Beijing) International Biotech Co., Ltd., China) $+$ water
\end{itemize}
Tween 20 and NF100 are conventional nonionic surfactants. Silwet 408 is an example of what is often referred to as a superspreader due to its very low surface tension and ability to spread over the leaf surface after impact \citep{Ananthapadmanabhan1990,Nikolov2002,Venzmer2011,Wang2013,Zhang2006}.

Each solution was sprayed onto leaves from three crops (cotton, rice and wheat) for both horizontal leaves ($0^\circ$) and leaves at a $45^\circ$ angle, to give 72 treatments. Plants were grown in a glasshouse for approximately one month prior to tests. Every treatment was repeated twice. The data from the three nozzle types was combined so that impaction results for each leaf type, spray mixture, and leaf angle combination were obtained from six separate leaves, each from a different plant.

The contact angle for each of the spray mixes on cotton leaves was initially measured by placing a 1 $\mu$L droplet with a micro syringe on a section of a cotton leaf, viewing side on through a microscope, manually measuring the height and base width of the droplet, and calculating the contact angle by the method in \citet{Mack1936}.  It was found that by the time the leaf section was positioned on the microscope, the microscope was focused and the measurements taken, droplets on wheat and rice leaves had flattened, resulting in lower than anticipated contact angle measurements.  To overcome this issue, the high speed video camera was then used to record images of the droplet as it was applied to the leaf.  The contact angle was determined from the resulting images by using the low-bond axisymmetric drop shape analysis (LBADSA) plug-in to ImageJ \citep{Stalder2010}.

To delineate impaction thresholds, a value for $K_\mathrm{crit}$, as described in \S~\ref{sec:shatter}, was calculated using contact angles of either $20\%$ or $50\%$ aqueous acetone solution via the method in \citet{Forster2010}. The $K_\mathrm{crit}$ value for each plant type is shown in Table \ref{tab:st_ca}.

Dynamic surface tension (DST) was measured with a Bubble pressure Tensiometer-BP2 (Kr\H{u}ss, Germany). It is well established that DST, rather than equilibrium surface tension, is better correlated with retention \citep{Anderson1989}. However, the most appropriate time scale for the DST measurement is not yet fully resolved. On one hand it may be argued that the droplet surface is instantaneously `fresh' at impact due to a significant deformation of the fluid free surface \citep{Anderson1989}, and so the most appropriate time scale may as early as $10$ ms. Others, however, argue that the DST measurement should correspond to the time to impact \citep{Anderson1989,Wirth1991,Stevens1993,Forster2005}. Since the most appropriate time scale for DST has not been resolved, we consider values at a surface age of both $10$ ms and $50$ ms. The former represents a fresh droplet surface, while the latter is the average time between atomization and impact. Since our DST data began at roughly $30$ ms, the $10$ ms estimate was obtained by extrapolation from a fitting of \citet{Hua1988}'s Equation (2) for the DST of a surfactant mixture. To arrive at the $50$ ms estimate, we assume the average velocity of a droplet from nozzle to target is roughly 10 m/s (all of the measured impact velocities in Figs~\ref{fig:horizontal} and \ref{fig:inclined} are less than 10 m/s, but drop velocities close to the nozzle are greater than 10 m/s, Wang et al., 2015), and the distance from nozzle to target is $0.5$ m.

For comparison, we have also considered DST at $900$ ms, which is essentially the equilibrium value for our three spray mixtures. The validity of using DST at these three ages is discussed in Sections \ref{sec:results_shatter} and \ref{sec:results_bounce}. The surface tension values for each spray mixture are shown in Table \ref{tab:st_ca}.
\begin{table*}
\caption{Surface tension and contact angles of the test spray mixtures on leaf surfaces. Standard deviation from 3 replicate measurements shown in brackets.}
\label{tab:st_ca}
\centering
\begin{tabular}{l l l l l l l}
\hline\noalign{\smallskip}
										& \multicolumn{3}{c}{Surface tension $\sigma$  (N/m)} & \multicolumn{3}{c}{Contact angle} \\
\noalign{\smallskip}\hline\noalign{\smallskip}
Mix 								&  $10$ ms        &  $50$ ms        &  $900$ ms				 & Cotton 	 						& Wheat 								& Rice \\
\noalign{\smallskip}\hline\noalign{\smallskip}
Water 							& 0.0716 (0.0001) & 0.0716 (0.0001) & 0.0716 (0.0001)  & 54$^{\circ}$ (10) 		& 132$^{\circ}$ (2) 		& 129$^{\circ}$ (6) \\
0.05\% NF100 				& 0.0640 (0.0002) & 0.0569 (0.0002) & 0.0482 (0.0002)  & 48$^{\circ}$ (3) 		& 110$^{\circ}$ (4) 		& 126$^{\circ}$ (3) \\
0.10\% Tween 20 		& 0.0457 (0.0001) & 0.0374 (0.0001) & 0.0320 (0.0001)  & 55$^{\circ}$ (7) 		& 116$^{\circ}$ (4) 		& 108$^{\circ}$ (4) \\
0.05\% Silwet 408		& 0.0514 (0.0002) & 0.0359 (0.0002) & 0.0194 (0.0002)  &  1$^{\circ}$* 				&  69$^{\circ}$ (9) 		&  73$^{\circ}$ (2) \\
\noalign{\smallskip}\hline\noalign{\smallskip}
									  & 								&  \multicolumn{2}{c}{$K_\mathrm{crit}$ (dimensionless)} & 150 & 69 & 74 \\
\noalign{\smallskip}\hline \\
 & & \multicolumn{3}{l}{*Complete spread so value of 1$^{\circ}$ used.} & &
\end{tabular}
\end{table*}


\section{Droplet impaction models}
\label{sec:models}

In \citet{Dorr2013a}, simple mathematical models were presented to predict the outcome of a droplet impacting a horizontal leaf surface from directly above (normal impaction). The three possible outcomes considered were shatter, bounce, and adhesion. The shatter criterion (based on Mundo et al. 1995) is summarised below in \S~\ref{sec:shatter}, and in \S~\ref{sec:bounce}, an extension to the bounce criterion (based on Mao et al. 1997) is provided to allow for different angles of the leaf surface and the droplet impact trajectory (oblique impaction). The key parameters describing droplet impaction are the impact velocity ($V$), droplet diameter before impact ($D$), fluid density ($\rho$), dynamic fluid viscosity ($\mu$), fluid-air surface tension ($\sigma$), and equilibrium contact angle ($\theta_\mathrm{e}$). Note that each spray mixture is assumed to have the density and viscosity of water ($\rho=1000$ kg/m$^3$ and $\mu=9.78\times10^{-4}$ Pa$\cdot$s); it is only surface tension that changes with the addition of an adjuvant (see Table \ref{tab:st_ca} for specific surface tension values).


\subsection{Shatter criterion}
\label{sec:shatter}

The process of droplet shatter can be quite variable, depending on the nature of the fluid and the surface being impacted. Types of shatter can be categorised as prompt, crown, and receding, to name a few \citep{Yarin2006,Marengo2011}. This wide variety, together with the fact that the physical processes occurring during shatter are not entirely understood, makes it a very difficult subject to accurately model and predict using simple models.

It is widely agreed that a prompt splash occurs when the inertial forces created by impact are sufficient to overcome the capillary forces which hold the fluid together \citep{Moreira2010}. This takes place during the spreading process, immediately after impaction. Several authors have developed criteria to predict shatter based on this theory, such as \citet{Mundo1995}, who proposed the relation
\begin{equation}
\label{eq:K}
K = \mathrm{We}_{\mathrm{n}}^{1/2}\mathrm{Re}_{\mathrm{n}}^{1/4} \,,
\end{equation}
where $\mathrm{We}_{\mathrm{n}}=\rho V_{\mathrm{n}}^2D/\sigma$ and $\mathrm{Re}_{\mathrm{n}}=\rho V_{\mathrm{n}}D/\mu$ are the dimensionless Weber and Reynolds numbers computed with the component of velocity normal to the impacted surface
\begin{equation}
\label{eq:Vn}
V_{\mathrm{n}} = V \sin\alpha \,.
\end{equation}
Here $\alpha$ is the angle between the leaf surface and the incoming trajectory of impact ($0<\alpha\leq 90^{\circ}$). Thus (\ref{eq:K}) is valid for both normal and oblique impactions.

A droplet is predicted to shatter on impact if
\begin{equation}
\label{eq:Kcrit}
K>K_\mathrm{crit}\,
\end{equation}
where $K_{\mathrm{crit}}$ is a critical value related to the properties of the surface being impacted. Mundo et al. found that $K_{\mathrm{crit}}=57.7$ correlated well with their data (which involved impactions of three different fluids onto smooth and rough surfaces, which rotated to mimic impact at an angle); however, this value could not be expected to translate to the data presented in this study. \citet{Mao1997}, for example, demonstrated that $K_{\mathrm{crit}}$ could be as large as $152$ for particular surfaces. Therefore criteria like (\ref{eq:Kcrit}) typically require that laborious adhesion and shatter experiments are performed in order to fit a suitable value of $K_{\mathrm{crit}}$ to any \emph{new} data set. To remove this limitation, \citet{Forster2010} devised a simple method for estimating $K_{\mathrm{crit}}$ based on contact angle measurements of standardised formulations. In this way, the value of $K_{\mathrm{crit}}$ adapts to reflect the wettability properties (including roughness effects) of the surface, while $K$ describes the fluid and droplet properties. Hence the novelty in this study is in the use of Forster et al.'s method to extend upon \citet{Mundo1995}'s criterion.

Note that as the angle of impact $\alpha$ decreases, $V_{\mathrm{n}}$ will in turn decrease, leading to a smaller calculated value of $K$. The implication of this trend is that shatter becomes less likely with a smaller impact angle.

It is worth noting that (\ref{eq:K}) may be equivalently written as $K=\mathrm{Oh}\,\mathrm{Re}_{\mathrm{n}}^{5/4}$, where $\mathrm{Oh}=\mu/(\rho g D)^{1/2}$ is the Ohnesorge number.  Further, (\ref{eq:Kcrit}) may be written as $\mathrm{Oh}> K_\mathrm{crit} \mathrm{Re}_{\mathrm{n}}^{-5/4}$.  Thus, as summarised by \cite{Rein2008}, this criterion belongs to a family of models predicting that shatter occurs when $\mathrm{Oh}> \alpha\, \mathrm{Re}_{\mathrm{n}}^{\beta}$, with $\alpha$ constant and $\beta=-5/4$. Other models in this family are due to \citet{Cossali1997} and \citet{Yarin1995}. In these works, the focus was on droplet impaction on a prewetted surface. Our experimental set up is such that droplets usually impact dry areas, although towards the end of the run there is an increased chance of impacting on portions of the leaf surface that had previously been wet by other droplets. Thus we do not adopt the models of \citet{Cossali1997} or \citet{Yarin1995}.

Mundo et al.'s shatter criterion, (\ref{eq:Kcrit}), provides a method for predicting whether splash will occur or not. However, it does not begin to address the secondary distribution of the spray, such as the number of secondary droplets produced in the splash, and their individual velocities and trajectories from the impaction site. Such post-shatter information is not considered in the current study, but relevant models have been published in \citet{Yoon2006} and \citet{Dorr2013a}.


\subsection{Bounce criterion}
\label{sec:bounce}

In the event that (\ref{eq:Kcrit}) predicts that shatter does not occur, a secondary criterion is required to differentiate between the possible subsequent outcomes of bounce and adhesion. Our bounce criterion hinges on work by \citet{Mao1997}, with a few modifications.

Mao et al.'s bounce model considers several key stages of the impaction process:
\begin{enumerate}[(a)]
\item Before impact. A spherical droplet of diameter $D$ is travelling at velocity $V$ towards a solid surface.
\item Maximum spread. After impact, the droplet spreads radially to a maximum diameter. A cylindrical disk geometry is assumed.
\item Maximum recoil. After maximum spread, the droplet recoils toward the point of impact.
\item Equilibrium. If the droplet does not bounce, it will settle to an equilibrium formation on the surface.
\end{enumerate}
In addition to these four, a fictitious stage is introduced for the purpose of predicting bounce:
\begin{enumerate}[(e)]
\item Bounce criterion. A spherical droplet momentarily at rest above the surface.
\end{enumerate}
Equating stages (a) and (b) leads to a cubic polynomial describing the maximum spread diameter of the droplet. It is assumed that, if bounce occurs, it takes place after stage (c), and instead of (d), in response to a sufficiently rapid recoil. A comparison between stage (c) and the fictitious stage (e) determines whether detachment from the surface (bounce) is the energetically favourable outcome after recoil. A successful bounce is indicated by a positive value of an `excess rebound energy':
\begin{equation}
\label{eq:Eere}
E_\mathrm{ERE}>0\,.
\end{equation}
If $E_\mathrm{ERE}<0$, then the droplet is predicted to adhere to the substrate.

\citet{Dorr2013a} expanded upon Mao et al.'s model by using $E_\mathrm{ERE}$ to calculate the exit velocity of the bouncing droplet. However, the model fundamentally describes the spread and bounce of a drop impinging a solid horizontal substrate from directly above (normal impaction). This is a major limitation in a spray scenario where leaf surfaces may have any orientation, and are impacted by spray droplets from various directions. Here we present a modification of the bounce criterion to account for oblique impaction.

The impact angle parameter $\alpha$, is used to quantify the `obliqueness' of an impact, with $\alpha = 90^{\circ}$ for a normal impaction and smaller $\alpha$ being more oblique. The orientation of the leaf surface itself is not taken into account explicitly because it is assumed that the magnitude of the inertial forces acting on the droplet during impaction, over the short timescale of spread and recoil, is much greater than the effect of gravity on the droplet. This is reasonable when considering typical spray-sized droplets of very small mass.

The modelling for oblique impact is kept as similar to the case of normal impaction as possible. Indeed, the criterion outlined here reduces to that in \citet{Dorr2013a} when $\alpha = 90^{\circ}$. While a cylindrical disk geometry is assumed by \citet{Mao1997} for a normal impaction, an elliptical disk geometry is now taken for the spreading drop, with the major axis in the impact direction. There is some probability of an oblique impaction leading to rivulet-type stretching, sliding, or rolling downhill (particularly for small $\alpha$ and highly hydrophobic surfaces; Antonini et al.~2014). However, it has been observed that a drop impacting an oblique surface can also exhibit the conventional spread and recoil behaviour that we assume here \citep{Liu2014}. Further, ellipses have been used in the literature to approximate the shape of oblique spreading patterns in a few applications \citep{Kang2006,Chen2005,Zen2010}.  Thus, we acknowledge that the use of an elliptical geometry and the assumptions behind it is an \emph{ad hoc} approach which meets our requirements of a simplistic criterion with realistic bounce trends.  There may be other simple alternatives that work better, especially in situations where $\alpha$ is expected to be small.

To quantify the amount of spread along the minor and major axes of the assumed ellipse, a few simple assumptions are made about the spreading process:
\begin{enumerate}[1.]
\item The sideways spread of the droplet (in the direction of the minor axis of the ellipse, perpendicular to the impaction direction) can be quantified using the normal component of impact velocity.
\end{enumerate}
Although the majority of spread will be in the impact direction, some spread will occur in the sideways direction as well. The sideways spread will decrease as $\alpha$ decreases, along with the normal component of impact velocity, $V_{\mathrm{n}}$.
\begin{enumerate}[2.]
\item The area of the droplet at maximum elliptical spread is equal to what it would be if impacting at $\alpha=90^{\circ}$ with the same impact velocity.
\end{enumerate}
The principle here is that the droplet should undergo roughly the same extent of deformation when impacting a surface at velocity $V$, regardless of whether the impaction is oblique or normal.
\begin{enumerate}[3.]
\item The retraction of the droplet after maximum spread will depend only on the major axis.
\end{enumerate}
A droplet bounces if there is sufficient surface energy to cause the droplet to recoil rapidly. If the spread is elliptical, the major axis will determine the likelihood of rapid recoil while the minor ellipse will have little contribution to the process.

The method employed for predicting oblique bounce is as follows. First, the impaction is treated as if it is normal ($\alpha=90^{\circ}$) \citep{Mao1997,Dorr2013a}. The maximum spread diameter $D_{\mathrm{normal}}$ is found via the cubic polynomial
\begin{align}
\label{eq:cubic}
\left[ \frac{1}{4}(1-\cos\theta_{\mathrm{e}}) + 0.2\frac{\mathrm{We}^{0.83}}{\mathrm{Re}^{0.33}} \right] \left( \frac{D_{\mathrm{normal}}}{D} \right)^3 \nonumber\\
- \left( \frac{\mathrm{We}}{12} + 1 \right) \left( \frac{D_{\mathrm{normal}}}{D} \right) + \frac{2}{3} = 0 \,,
\end{align}
which can be solved exactly for $D_{\mathrm{normal}}$. This corresponds to Equation (17) of \citep{Mao1997}, and makes use of a stagnation point flow analogy to empirically quantify viscous dissipation during the spreading process.

If a real solution to (\ref{eq:cubic}) does not exist, or $D_{\mathrm{normal}}$ is calculated to be less than $D$, we set $D_{\mathrm{normal}} = D$ as a minimum bound. Once $D_{\mathrm{normal}}$ is known, the second assumption implies that the spread area of the droplet in the oblique impaction is
\begin{equation}
\label{eq:area}
\text{Area} = \frac{\pi}{4}D_{\mathrm{normal}}^2 \,.
\end{equation}
Using the first assumption, the extent of sideways spread of the elliptical droplet is calculated with the normal component of the velocity (Equation (\ref{eq:Vn})), leading to another cubic equation for maximum spread:
\begin{align}
\label{eq:cubic2}
\left[ \frac{1}{4}(1-\cos\theta_{\mathrm{e}}) + 0.2\frac{\mathrm{We}_{\mathrm{n}}^{0.83}}{\mathrm{Re}_{\mathrm{n}}^{0.33}} \right] \left( \frac{D_{\mathrm{minor}}}{D} \right)^3 \nonumber\\
- \left( \frac{\mathrm{We}_{\mathrm{n}}}{12} + 1 \right) \left( \frac{D_{\mathrm{minor}}}{D} \right) + \frac{2}{3} = 0 \,,
\end{align}
to be solved for $D_{\mathrm{minor}}$ under the condition. The notation $\mathrm{We}_{\mathrm{n}}$ and $\mathrm{Re}_{\mathrm{n}}$ denotes the Weber and Reynolds numbers computed with $V_{\mathrm{n}}$ rather than $V$. If $D_{\mathrm{minor}}<D$, then we set $D_{\mathrm{minor}}=D$ as a correction.

Now, the area of an ellipse is given by
\begin{equation}
\label{eq:area2}
\mathrm{Area}=\frac{\pi}{4} D_{\mathrm{minor}} D_{\mathrm{major}} \,,
\end{equation}
and so the second assumption coupled with (\ref{eq:area}) allows the calculation of $D_{\mathrm{major}}$ as
\begin{equation}
\label{eq:Dmajor}
D_{\mathrm{major}} = \frac{D_{\mathrm{normal}}^2}{D_{\mathrm{minor}}} \,.
\end{equation}
Using the third assumption, the excess rebound energy of the droplet is calculated by
\begin{align}
\label{eq:Eere2} E_{\mathrm{ERE}} = &\left[ \frac{\pi}{4}D_{\mathrm{major}}^2(1-\cos\theta_{\mathrm{e}}) + \frac{2}{3}\pi\frac{D^3}{D_{\mathrm{major}}} \right] \sigma \nonumber\\
&- 0.12\pi D^2\sigma \left( \frac{D_{\mathrm{major}}}{D} \right)^{2.3} (1-\cos\theta_{\mathrm{e}})^{0.63} \nonumber\\
&- \pi\sigma D^2 \,.
\end{align}
Bounce occurs if the criterion (\ref{eq:Eere}) is satisfied (otherwise adhesion is predicted). Equations (\ref{eq:cubic})-(\ref{eq:Eere2}) incorporate properties of the fluid (density, viscosity, and surface tension) and droplet (velocity and diameter) through the Weber and Reynolds numbers. Meanwhile, the equilibrium contact angle $\theta_\mathrm{e}$ describes the energetics of the liquid-solid-gas system. \citet{Mao1997} found that $\theta_\mathrm{e}$ was sufficient to account for the effects of surface roughness without explicitly including a roughness parameter. The exit velocity of the droplet is
\begin{equation}
\label{eq:v_exit}
V_{\mathrm{exit}} = \sqrt{ \frac{12E_{\mathrm{ERE}}}{\pi\rho D^3} } \,.
\end{equation}
The direction that the droplet bounces is assumed to simply be a mirror of its incoming trajectory. The validity of this assumption is discussed in \S~\ref{sec:results}.

The impaction models described here are best visualised by a bounce or shatter boundary. For a given plant and spray formulation, one could run the bounce criterion for an appropriate range of $V$ and $D$, and then on the $V$ versus $D$ plane delineate regions of predicted adhesion and bounce with a curve representing the boundary between the two outcomes. This boundary corresponds to $E_\mathrm{ERE}=0$. Similarly, $K=K_\mathrm{crit}$ leads to a shatter boundary which may be plotted on the same plane. The interpretation of the bounce (or shatter) boundary is that a spray droplet with initial velocity and diameter falling above the curve results in a positive prediction for bounce (or shatter). Recall that bounce may only occur if there was no shatter during the droplet's spreading process, and so shatter always takes precedence over bounce in the impaction model.


\section{Results}
\label{sec:results}

\begin{figure*}
\centering
\includegraphics[width=1.8\columnwidth]{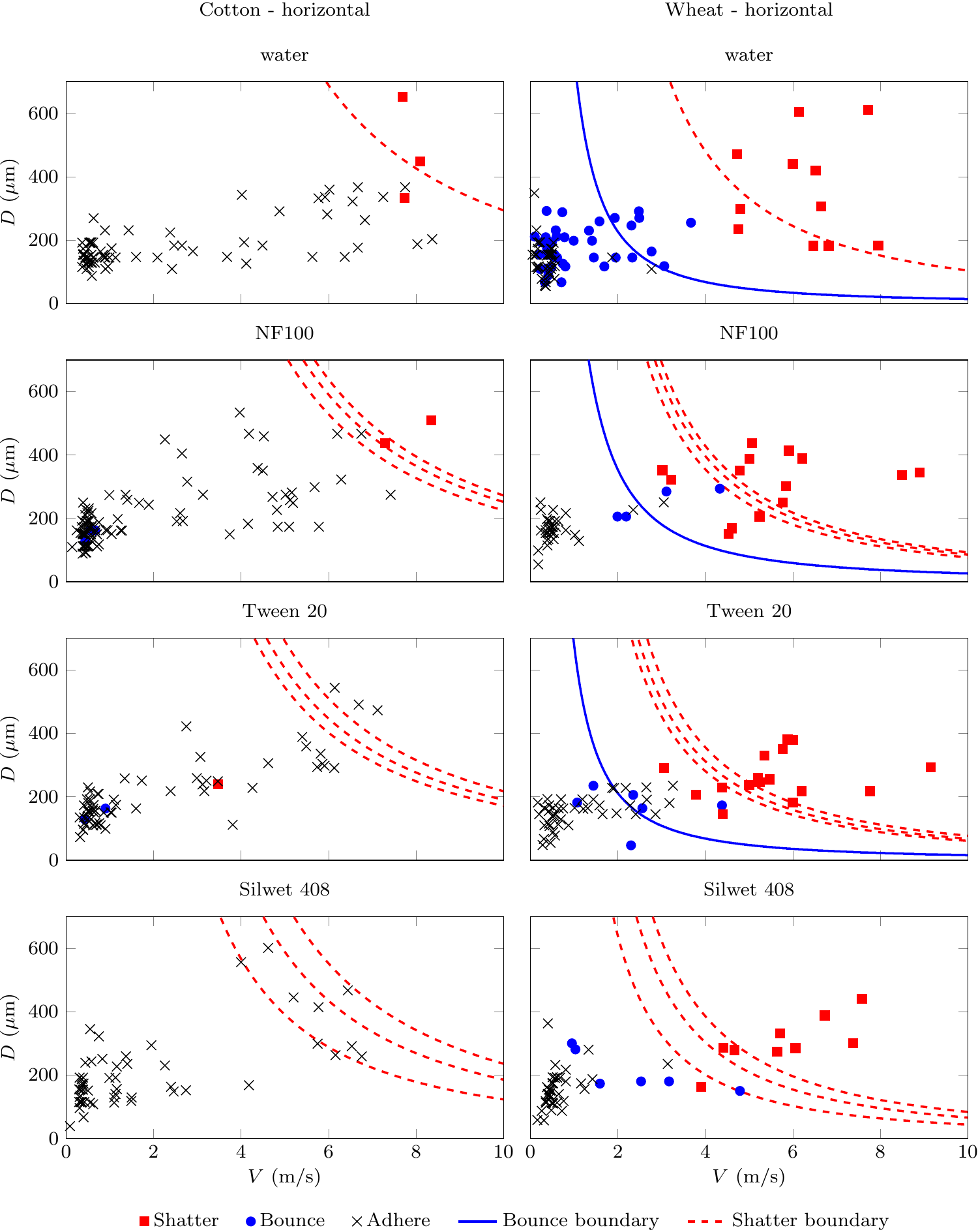}
\caption{Impaction observations for horizontal wheat and cotton leaves and a variety of formulations. The corresponding predictions are illustrated with bounce ($E_\mathrm{ERE}=0$) and shatter ($K=K_\mathrm{crit}$) curves. The top shatter curve in each subplot corresponds to DST at a surface age of $10$ ms, the middle curve to $50$ ms, and the bottom curve to $900$ ms. The bounce curves are computed with DST at $50$ ms.}
\label{fig:horizontal}
\end{figure*}

\begin{figure*}
\centering
\includegraphics[width=1.8\columnwidth]{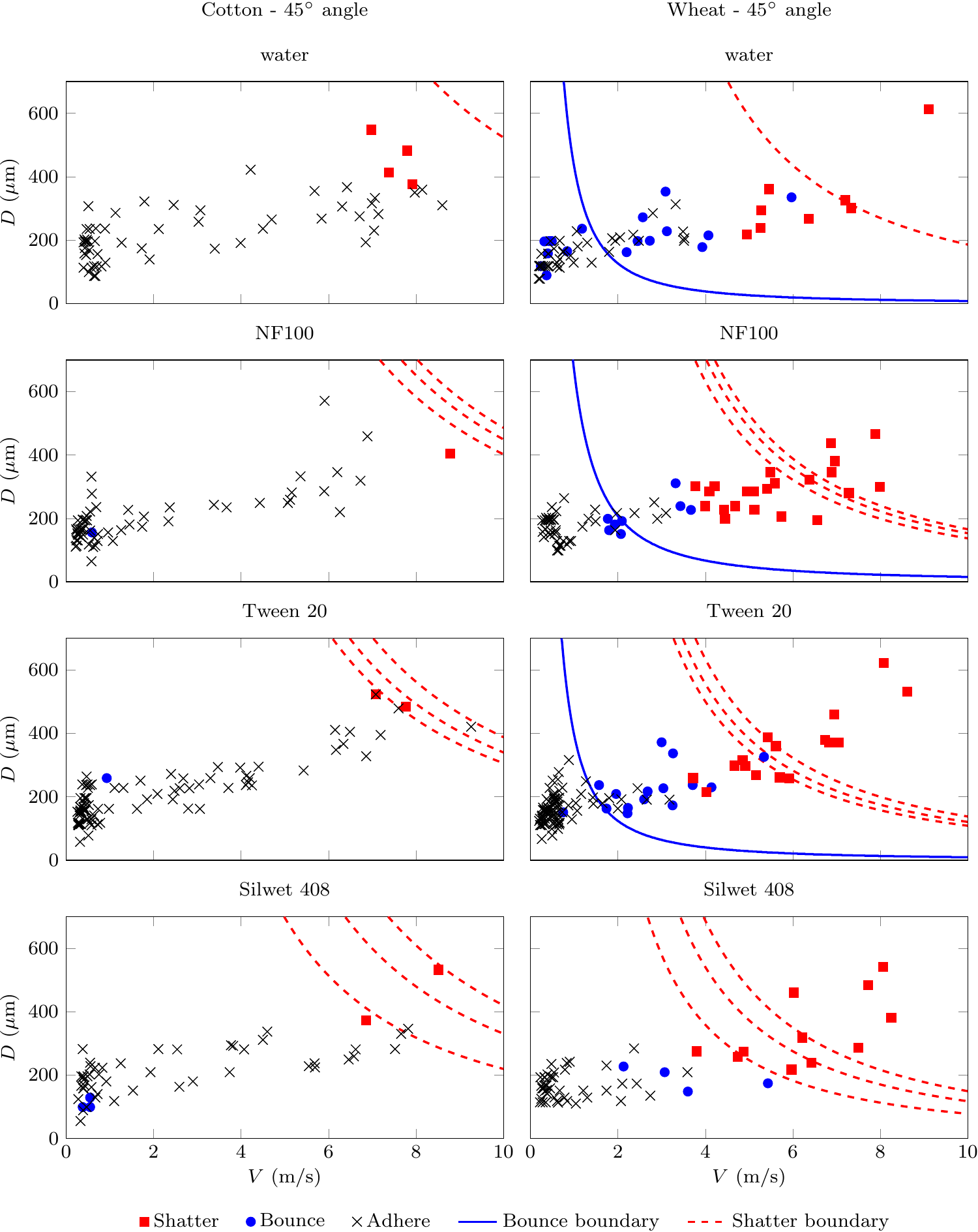}
\caption{Impaction observations for wheat and cotton leaves inclined at $45^\circ$, and a variety of formulations. The corresponding predictions are illustrated with bounce ($E_\mathrm{ERE}=0$) and shatter ($K=K_\mathrm{crit}$) curves. The top shatter curve in each subplot corresponds to DST at a surface age of $10$ ms, the middle curve to $50$ ms, and the bottom curve to $900$ ms. The bounce curves are computed with DST at $50$ ms.}
\label{fig:inclined}
\end{figure*}

Experimental impact outcomes for droplets on cotton and wheat leaves are shown in Fig~\ref{fig:horizontal} for horizontal leaf surfaces, and Fig~\ref{fig:inclined} for leaf surfaces inclined at an angle of 45$^{\circ}$. For each leaf type, the results for the four spray mixtures (water, Tween 20, Silwet 408 and NF 100) are shown separately. Each of the subplots combines data from the three nozzle types, and on average 75 impact outcomes, on six individual leaves, from six different plants, are recorded in each.  Additional data for rice is provided in the online supplementary material.

Each data point in Figs~\ref{fig:horizontal} and \ref{fig:inclined} represents a single droplet impaction, plotted according to the measured droplet diameter and velocity on impact. Three symbols are used to indicate the observed impaction outcomes: a (red) square for droplet shatter on impact, a (blue) circle for droplet bounce on impact, and a (black) cross for droplet adhesion on impact. All three outcomes are well represented in the results for wheat leaves, while the results for cotton leaves show a majority of adhesion observations with relatively few shatter and bounce observations. For all leaf and spray mixture types, there is not always a clear grouping of the three types of impaction outcome. This is particularly true for bounce and adhesion observations, where the data are often mixed in the lower left region of the plots (corresponding to smaller diameters and velocities). The shatter observations, on the other hand, are generally grouped in the upper right region (corresponding to larger diameters and velocities) with little overlap with bounce and adhesion observations. It is interesting to note that there is not a clear difference in the distribution of shatter, bounce and adhesion results between the horizontal and $45^\circ$ leaves.

For comparison of the experimental data with impaction criteria, shatter boundaries ($K=K_\mathrm{crit}$, from the model in \S~\ref{sec:shatter}) are indicated by a (red) dashed line, and bounce boundaries ($E_\mathrm{ERE}=0$, as described in \S~\ref{sec:bounce}) by a (blue) solid line. Recall that shatter is predicted when a droplet's impact diameter and velocity fall above the shatter boundary, so that (\ref{eq:Kcrit}) is satisfied. On the other hand, bounce is predicted when the droplet's attributes place it \emph{between} the bounce and shatter boundaries, because shatter takes precedence over bounce. In those plots within Figs \ref{fig:horizontal} and \ref{fig:inclined} where only a shatter boundary is depicted, the bounce boundary has been omitted because it fell above the shatter boundary. Droplet adhesion is predicted when the diameter and velocity information falls below or to the left of a bounce boundary, or below and to the left of the shatter boundary for cases without a bounce boundary. Three shatter boundaries are shown in each subplot of Figs~\ref{fig:horizontal} and \ref{fig:inclined}; these correspond to DST measurements at $10$ ms (top shatter boundary), $50$ ms (middle shatter boundary), and $900$ ms (bottom shatter boundary). Bounce boundaries are shown for a DST measurement at $50$ ms only. This is for two reasons. First, a change in DST made little visible difference to the position of the bounce boundary, and so does not affect interpretation of the results. Second, the most appropriate DST measurement is even more uncertain for the bounce process than it is for shatter, due to the larger timescale over which it occurs and the potential for DST to vary significantly over that time.

Note that there are a few important trends in the positions of the shatter and bounce boundaries in Figs~\ref{fig:horizontal} and \ref{fig:inclined}. First, the position of the shatter boundary lowers as the surface tension of the spray mix is decreased. This is true both between subplots (when comparing spray formulations), and within subplots (when comparing DST measurements). This means that the probability of shatter increases with decreasing surface tension, because the droplets have a lower surface energy and are therefore more prone to break-up. Second, the position of the shatter boundary rises as the leaf surface is tilted from the horizontal to $45^\circ$. This means that shatter becomes less probable in an oblique impaction. Finally, the bounce boundary slightly lowers as the leaf is tilted to $45^\circ$, indicating a higher probability of bounce in an oblique impaction.

\begin{figure*}
\includegraphics[width=\textwidth]{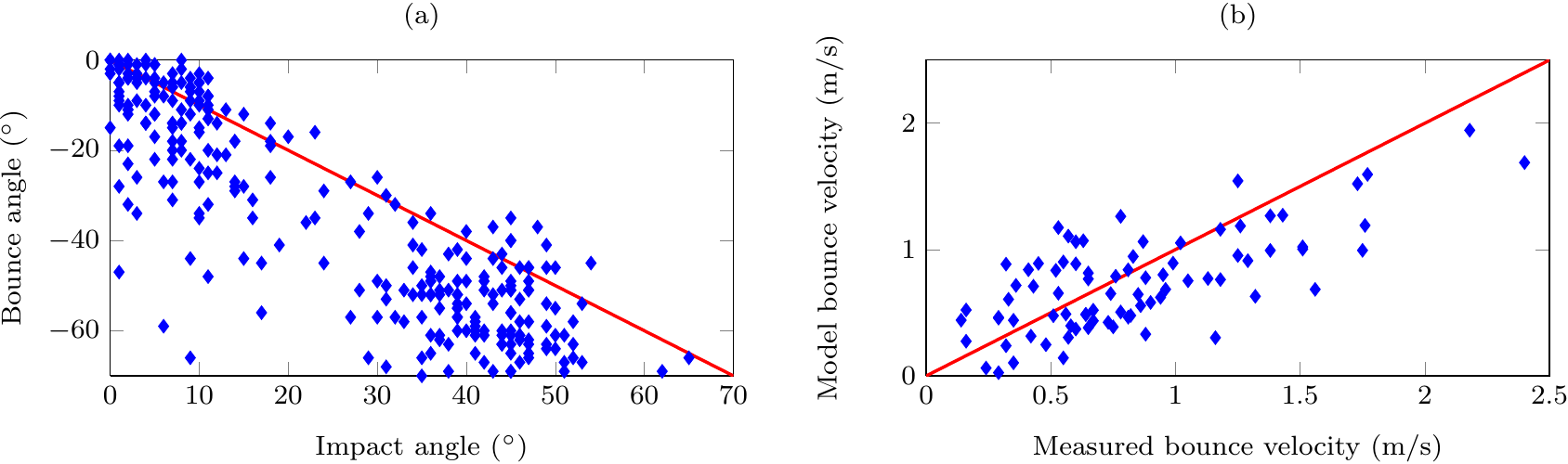}
\caption{(a) Measured impact angle and bounce angle for all droplets that were observed to bounce.  The (red) line shows the expected angle if bounce were to occur as a mirror image to the incoming trajectory, about the surface normal. (b) Measured velocity of droplet after bouncing off a surface compared to the predicted velocity.  The (red) line indicates where measured velocity would the same as the predicted velocity.}
\label{fig:bounce_angle_velocity}
\end{figure*}

In Fig~\ref{fig:bounce_angle_velocity}(a), the observed angle of bounce, as measured from the nominal surface normal, is contrasted with the angle of impact. The (red) line running diagonally through the plot indicates the expected bounce angle if rebound was to occur as a mirror image of the incoming trajectory. The distribution of the majority of data points below the line indicates that the bounce angle tends to be greater than the impact angle, implying that the assumption made in \S~\ref{sec:bounce} about these angles being equal is not strictly applicable.  This trend can be attributed to energy loss in the impact process (spread and retraction of the droplets).  It should be noted that while these results are relative to the nominal surface normal (normal to either a horizontal or $45^\circ$ surface), the leaf surfaces have irregularities that would cause the actual surface normal at the point of impact to vary from the nominal surface normal. This difference would contribute to the scatter in the data, and may account for the cases where the bounce angle is less than the impact angle.

Figure~\ref{fig:bounce_angle_velocity}(b) compares the measured velocity of bouncing droplets to the predicted velocity calculated by Equation (\ref{eq:v_exit}). The (red) line illustrates where data points would fall if there was perfect agreement between the experiments and prediction. There is a slight tendency for droplets with higher observed bounce velocities ($>1$ m/s) to fall below the line, indicating that bounce velocity may be underestimated by the model. The overall agreement is quite good, however, with the scatter of data points generally centered about the line.


\section{Discussion}
\label{sec:discussion}

While there exist a variety of sophisticated mathematical models that are designed to simulate the entire droplet impaction process, the majority of these are computationally intensive even for a single droplet.  On the other hand, we are motivated by large scale applications where thousands of drops may be present at the same time, such as in multicomponent simulations of agricultural spray scenarios.  Thus we have a need for simple, easy-to-compute models and, as such, have focused on algebraic energy balance criteria and an \textit{ad hoc} type approach.  While there is scatter present in the experimental data presented here, and distinct grouping of different impact outcomes (adhesion, shatter or bounce) is not always observed in Figs~\ref{fig:horizontal} and \ref{fig:inclined}, we find the overall predictions of the simple energy balance criteria compare well to the measurements.

Most of the experimental studies on droplet impaction to date have used uniform synthetic (dry) targets and mono-sized droplets.  However, in this work, we have used real leaves, with all their inherent natural variability (for example, hairs, trichomes, stomata, veins, wax structures), and commercial agricultural spray nozzles that result in a range of droplet velocities, sizes and trajectories being produced \citep{Dorr2013}. The nature of the spray experiments introduces the further complication that, while the leaf surfaces are initially dry, they do become partially wet throughout the run as drops accumulate on the surface. There are some measurement errors associated with the camera field of view and using a single camera to record a 3D process \citep{Dong2013,Dong2014}, although we have tried to minimise these by only selecting measurement of drops that remain in good focus throughout the impaction process.


\subsection{Shatter}
\label{sec:results_shatter}

The shatter boundary tends to work reasonably well for the situations tested, although there are some discrepancies where most shatter is observed below the predicted shatter line (see water on $45^\circ$ cotton and water and NF 100 on $45^\circ$ wheat in Fig~\ref{fig:inclined}). There is also little difference between shatter boundaries for the three DST values, apart from in the case of the superspreader mix where the boundaries are more spread apart. Therefore it is difficult to make any conclusions about the most appropriate timescale for DST with respect to shatter. The occurrence of droplet shatter increases with diameter and velocity, as evidenced by the concentration of shatter observations in the upper-right of the plots. This trend is more obvious on the difficult to wet wheat leaves than the easy to wet cotton leaves where shatter observations were more scarce.

In theory we might expect to find that the number of shatter observations increases as the surface tension of the spray mix is decreased, and that a higher impact velocity would be required for shatter as the surface tension is increased. This is due to the hydrodynamic principle that the inertial forces of impact must overcome the capillary effects, which are weaker as the surface tension decreases. There is perhaps some evidence in Figs~\ref{fig:horizontal} and \ref{fig:inclined} that there are fewer shatter observations for water on wheat than the other spray mixes, and that those observations generally occur at higher velocities. However, more data is required for confirmation.

As stated above, there is a tendency for some drops to shatter when their velocity and diameter fall below the shatter boundary. Interestingly, these discrepancies tend to be present for the oblique impactions (Fig~\ref{fig:inclined}) more than the normal impactions (Fig~\ref{fig:horizontal}). It is likely that at least some of these droplets hit a previously wet area on the leaf and are hence more prone to shatter \citep{VanderWal2006}, which is supported by the fact that these cases often appear later in the run where there is a higher chance of impacting a previously wet area. As the camera view is looking side on to the leaf surface, it is often not possible to determine if the impact area has been wet by previous droplet impacts or not.


\subsection{Bounce}
\label{sec:results_bounce}

Due to the overlap of bounce and adhesion results in Figs~\ref{fig:horizontal} and \ref{fig:inclined}, there is not always a distinct region of bounce only observations in the data as there was for shatter, but the expected trends are still present overall. For example, the bounce criterion predicts that droplets will never bounce on the more easily wettable cotton, or with superspreader spray mixtures such as Silwet. In the data, only a few droplets were observed to bounce for these cases, particularly when compared to the number of observations for bouncing of water droplets.

Many bounce observations can be seen below the bounce boundaries in Figs~\ref{fig:horizontal} and \ref{fig:inclined}, implying that the probability of bounce is higher than was predicted by the bounce criterion. This is particularly clear in the plot for water on a horizontal wheat leaf in Fig~\ref{fig:horizontal}, where many bounce outcomes were observed for small drops of low impact velocity, but not predicted by the bounce criterion due to low kinetic energy.

This discrepancy between the predictions and data could be due to the fundamental assumptions of the bounce criterion, adapted from \citet{Mao1997}. Mao et al.~acknowledge that the precision of the model decreases significantly for low impact velocities ($V<1$ m/s) due to the assumption of a cylindrical disk spreading geometry, and a stagnation point flow analogy for viscous dissipation.

One consideration may arise from the bounce observations common to the context of superhydrophobic surfaces. In such studies \citep{Richard2000,Okumura2003,Liu2014,Moevius2014}, the substrate is typically textured to induce a Cassie-Baxter wetting regime, minimising both contact angle hysteresis and contact area between fluid and substrate on impact. As a result, drops impacting such surfaces have a greatly enhanced tendency to bounce. \citet{Richard2000} observed that bouncing drops with $\mathrm{We}\ll1$ (kinetic energy significantly smaller than surface energy) had very little shape deformation on impact, and a high restitution coefficient ($e = V_\mathrm{exit}/V$) due to efficient transfer of kinetic energy to surface energy. Viscous dissipation within the fluid was shown to be negligible. The largest contact angle considered in this study is $132^\circ$, so our surfaces are not considered superhydrophobic, but these results illustrate why drops with low kinetic energy may not be adequately described by the current model. A different geometry for the impacting drop (such as an oblate spheroid, rather than a cylinder) may be required for this regime, as well as a reevaluation of the dissipation term.

Another significant assumption of the bounce criterion is that a drop will make contact with the leaf surface on impact. In reality, there is an extremely small air layer between the drop and substrate which must first be drained before contact between the two can occur \citep{Liu2015,Ruiter2015}. Therefore the discrepancy between the predictions and data could be due, at least in part, to complex air layer dynamics that are not taken into account in the bounce equations presented in \S~\ref{sec:bounce}.

\citet{Kolinski2014a} demonstrated that, on even a perfectly hydrophilic substrate, the air layer can prevent contact and cause a drop to rebound. However, this was only true for an atomically smooth substrate, and it was shown that a drop would inevitably make contact with a surface with defects larger than the thickness of the film (as would be true for most surfaces, particularly natural surfaces such as leaves). Air layer effects have also been show to enhance the rebound of drops impacting on liquid. For example, \citet{Couder2005} demonstrated bouncing on the surface of a vertically vibrated liquid bath. Of more relevance to our study, \citet{Gilet2012} described the presence of an air layer aiding the rebound of small, slow-moving drops, which were impinging a surface prewetted with a viscous fluid.  They observed that a compressible air layer between the drop and underlying film could act as a cushion, promoting bounce as long as the layer was not penetrated. On the other hand, faster drops could break through the air layer, make contact with the wetted surface, and adhere. These observations could represent impactions occurring towards the end of our experiments, when there is a small possibility of droplets hitting a previously wet area.

\begin{figure}
\centering
\includegraphics[width=\columnwidth]{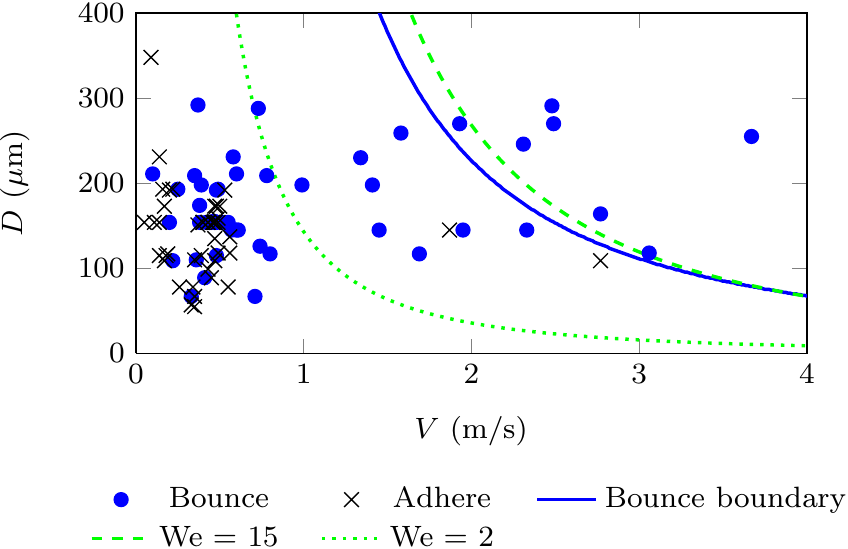}
\caption{Magnified chart of water droplets impacting a horizontal wheat leaf.  The predicted bounce boundary presented in this paper is indicated, along with curves corresponding to $\mathrm{We}=2, 15$.}
\label{fig:bounce_We}
\end{figure}

\citet{Gilet2012} reported that enhanced bounce due to the air layer effect occurred for $\mathrm{We}<2$. They found that when $\mathrm{We}>15$, the air layer breaks and a total merger occurs between the droplet and the underlying film. Between $\mathrm{We}=2$ and 15, a complex series of intermediate regimes occur.  For context,  curves indicating $\mathrm{We}=2$ and $\mathrm{We}=15$ have been added to the impact outcome plot for water on a horizontal wheat leaf (refer to Fig~\ref{fig:horizontal}) in Fig~\ref{fig:bounce_We}. In Gilet and Bush's observations, droplets falling above the $\mathrm{We}=15$ curve were not susceptible to the air layer effect, and those below the $\mathrm{We}=2$ curve were.

We have presented some of many possible explanations for the discrepancy between observed and predicted frequencies of bounce. Other potential causes could be the presence of dust particles on the leaves or irregularities in the leaf surface structure, neither of which are accounted for in the bounce criterion.

In terms of total spray retention on whole plants, this extra incidence of bounce may not have a significant influence, since these small droplets only account for a very small proportion of the total volume of the spray, and they are typically moving at low velocities so they are often recaptured by the same leaf anyway.

As with shatter discussed above, the value of dynamic surface tension we use in our model (measured at 50 ms) may not be the most appropriate to describe the bouncing process. Since bounce occurs over a considerably longer timescale than shatter, the concentration of surfactant molecules at the fluid surface can vary in a complex manner during the processes of spreading and recoiling, which is not yet completely understood. Therefore, different values of $\sigma$ (and consequently, $\theta_\mathrm{e}$) may need to be used for each term of Equation (\ref{eq:Eere2}) to reflect the true dynamic nature of the surface tension.


\subsection{Surface wettability, roughness and texture}

The properties contributing to the wettability of a leaf can be divided into two categories: physical and chemical interactions between the leaf surface and the droplet solution \citep{Holloway1970}. Most studies do not attempt to isolate and quantify each of these properties, but rather indirectly infer relative values by quantitatively measuring behaviour that is allied to wettability, such as contact angles \citep{Nairn2011} (as we have done in this paper).

A leaf may be difficult-to-wet (high contact angle of water, 20\% or 50\% acetone) even though it may have a chemically neutral or even relatively polar surface, due to the surface being extremely rough (as opposed to being hydrophobic) \citep{Nairn2011,Nairn2015}.

\begin{figure}
\centering
\includegraphics[width=0.7\columnwidth]{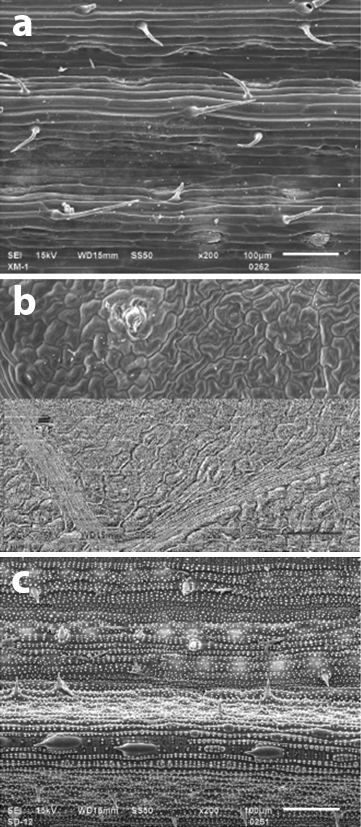}
\caption{Scanning electron microscope (SEM) photographs at 200-fold magnification of leaves for three crops (a) wheat, (b) cotton, (c) rice.}
\label{fig:surf}
\end{figure}

In order to demonstrate the variability of the leaf surfaces from each of the three crops at a small scale, scanning electron microscope (SEM) photographs at 200-fold magnification \citep{Yang2012} are presented in Fig~\ref{fig:surf}.  These images show: the surface of a wheat leaf is smooth and covered by a wax layer with hooked shape (unciform) and needle shaped (acicular) hair on the surface, but no papilla (Fig \ref{fig:surf}a); the surface of a cotton leaf has no hair nor papilla (Fig \ref{fig:surf}b); and the surface of rice leaf is rough with unciform hairs and also papilla (Fig \ref{fig:surf}c).

\subsection{Leaf elasticity}

The impaction models presented in this study assume a stationary, rigid leaf surface. This is in line with the experimental observations gained from high-speed video imaging, in which the spray droplets were very small relative to the leaf surface, the leaf was fixed on a board, and there was no evidence of deformation of the leaf surfaces on impact. It is possible, however, that the impaction of large drops, the bombardment of many small spray drops, or the presence of wind could cause significant movement of leaves in a real spray scenario. Further, particular leaves may have compliant surfaces, especially when impacted by a drop with significant kinetic energy (size or velocity).

To incorporate leaf movement in the current bounce and shatter criteria, a straightforward approach may be to replace impact velocity with the net impact velocity (sum of the droplet and leaf velocities). However, \citet{Lee2004} demonstrate for the case of droplet rebound that it is the motion of the target surface and surrounding air which influences the impact outcome, rather than the net change in velocity. Therefore air motion may need to be considered, at least in the spreading stage where it is most influential.

Elasticity of the leaf surface may be accounted for with a small reduction in droplet kinetic energy, representing the kinetic energy absorbed by the surface during impact.  \citet{Pepper2008} showed that substrate elasticity can suppress shatter for this reason, but that these effects cannot be described by a traditional energy balance approach due to the timescale over which shatter occurs. Similarly, \citet{Mangili2012} and \citet{Alizadeh2013} considered drop impaction on soft polydimethysiloxane (PDMS) substrates, and demonstrated an absorbance of kinetic energy due to substrate deformation, which resulted in a significant reduction of droplet recoil. This suggests a lower available energy for rebound on compliant surfaces. In agreement with these studies, \citet{Cho2013} saw that their fibrous elastic membrane could suppress both shatter and bounce.

Complicating these analyses are the findings of \citet{Chen2010} and \citet{Chen2011}, which show that an air film may form beneath a drop impacting a soft PDMS substrate. As long as the drop velocity is within a critical lower and upper limit, air entrapment occurs between the droplet and surface due to the shear-thinning property of elastomeric PDMS. As a result, droplet bounce was observed on soft PDMS even when it did not occur for any impact velocity on rigid PDMS. These results, along with the discussion of air films in the previous section, illustrate a need for a better understanding of air layer effects on droplet impaction.


\subsection{Dispersal of pathogens by rain}

While the data presented in this paper were generated for typical pesticide applications, the principles are relevant to the impaction of rain droplets onto leaf surfaces and the subsequent dispersal of pathogens between plants.  The main differences between these two applications are that: 1) rain droplets are purely water, whereas typical spray mixtures for pesticide application contain additions that alter characteristics such as surface tension; and 2) for pesticide applications, the leaf surfaces are mostly dry and the volumes used are usually such that the surface does not become fully wet, while impaction during rain events are often onto fully wet surfaces.  It would be expected that shattering of rain droplets is enhanced through the combination of wet surfaces and larger droplets while shattering of pesticide drops is enhanced through lowering of surface tension. The bounce and shatter criteria developed in this paper assume a dry surface and additional work is required to characterise the influence of a wet surface on droplet impaction outcomes. The post-impact behaviour of droplets for both rain dispersal and pesticides needs to be characterised through further research to enable distribution of either pathogens or pesticide through a plant canopy to be determined.


\section{Conclusions}
\label{sec:conclusions}

The application of agricultural chemicals via spraying, and spreading of pathogens by rain splash, are complicated processes and there are many inherent variations.  These include: the plant shape and structure with resulting variation in leaf angles; leaf microstructure that can contain, for example, hairs, trichomes, stomata, veins, and wax structures; variable droplet properties (in particular size, velocity, trajectories, density); variable formulation properties (e.g. surface tension and viscosity); variable droplet impact conditions (due to the increased possibility of impaction on a previously wet area as the run progresses); and variable airflow due to micrometeorology and turbulence resulting from the spray application.  The experimental technique used in this study incorporated many of these variables, such as the use of real leaves with all their inherent natural variability, and commercial agricultural spray nozzles resulting in a range of droplet velocities, sizes and trajectories being produced.  While there is scatter in the experimental measurements largely due to these uncontrolled factors, the results do show broad agreement with predictions.  In particular:
\begin{itemize}
\item	shatter of droplets increased with increasing size and velocity of droplets at impact;
\item	shatter of droplets increased as the surface become harder to wet (shatter was more obvious on the hard to wet wheat leaves than the easy to wet cotton leaves);
\item	for easy to wet surfaces or superspreader mixtures, the droplets mostly either adhered or shattered on impact, with very few droplets bouncing;
\item	bouncing of droplets was most pronounced on hard to wet surfaces (such as wheat) with high surface tension spray mixtures (such as water).
\end{itemize}
Droplets which had a diameter of 200 $\mu$m or smaller and were travelling at less than 1-2 m/s mostly adhered on impact, with a few exceptions, particularly in the case of water droplets on wheat. On wheat, many of these smaller, slow moving droplets were observed to bounce.  Shatter became predominant as the velocity and size of droplets was increased, without exception.  Addition of adjuvants (surfactants) which lower surface tension can assist with initial adhesion, but will also lead to more shatter. By combining modeling with experimental observations, a better insight into the processes occurring with different spray conditions has been achieved.


\begin{acknowledgement}

This work is funded by the Australian Research Council through the ARC Linkage Project LP100200476, its industry partners (Syngenta, Dow AgroSciences, Crop-lands/NuFarm, Bill Gordon Consulting and Plant Protection Chemistry NZ Ltd), and China Bureau of Foreign Experts through the foreign experts project of ``Development of loss reducing techniques for chemicals application of field crop in China'' (Project No: 2012Z022).  The authors are particularly grateful to Jerzy (George) Zabkiewicz for many helpful discussions.

\end{acknowledgement}




\begin{thebibliography}{}

\bibitem[Alizadeh et al.(2013)]{Alizadeh2013} Alizadeh A, Bahadur V, Shang W, Zhu Y, Buckley D, Dhinojwala A (2013) Influence of substrate elasticity on droplet impact dynamics. Langmuir 29:4520-4524

\bibitem[Ananthapadmanabhan et al.(1990)]{Ananthapadmanabhan1990} Ananthapadmanabhan KP, Goddard ED, Chandar P (1990) A study of the solution, interfacial and wetting properties of silicone surfactants. Colloids and Surfaces 44:281-297

\bibitem[Anderson and Hall(1989)]{Anderson1989} Anderson NH, Hall DJ (1989) The role of dynamic surface tension in the retention of surfactant sprays on pea plants. Adjuvant and Agrochemicals 2:51-62

\bibitem[Antonini et al.(2014)]{Antonini2014} Antonini C, Villa F, Marengo M (2014) Oblique impacts of water drops onto hydrophobic and superhydrophobic surfaces: outcomes, timing, and rebound maps. Experiments in Fluids 55

\bibitem[Attan\'{e} et al.(2007)]{Attane2007} Attan\`{e} P, Girard F, Morin V (2007) An energy balance approach of the dynamics of drop impact on a solid surface. Physics of Fluids 19:012101

\bibitem[Butler-Ellis et al.(2002)]{ButlerEllis2002} Butler-Ellis MC, Swan T, Miller PCH, Waddelow S, Bradley A, Tuck CR (2002) Design factors affecting spray characteristics and drift performance of air induction nozzles. Biosystems Engineering 82:289-296

\bibitem[Calonnec et al.(2013)]{Calonnec2013} Calonnec A, Burie J-B, Langlais M, Guyader S, Saint-Jean A, Sache I, Tivoli B (2013) European Journal of Plant Pathology 135:479-497

\bibitem[Chen and Wang(2005)]{Chen2005} Chen RH, Wang HW (2005) Effects of tangential speed on low-normal-speed liquid drop impact on a non-wettable solid surface. Experiments in Fluids 39:754-760

\bibitem[Chen and Li(2010)]{Chen2010} Chen L, Li Z (2010) Bouncing droplets on nonsuperhydrophobic surfaces. Physical Review E 82:016308

\bibitem[Chen et al.(2011)]{Chen2011} Chen L, Wu J, Li Z, Yao S (2011) Evolution of entrapped air under bouncing droplets on viscoelastic surfaces. Colloids and Surfaces A 384:726-732

\bibitem[Cho et al.(2013)]{Cho2013} Cho SJ, Nam H, Ryu H, Lim G (2013) A rubberlike stretchable fibrous membrane with anti-wettability and gas breathability. Advanced Functional Materials 23:5577-5584

\bibitem[Cossali et al.(1997)]{Cossali1997} Cossali GE, Coghe A, Marengo M (1997) The impact of a single drop on a wetted surface.  Experiments in Fluids 22:463-472

\bibitem[Couder et al.(2005)]{Couder2005} Couder Y, Fort E, Gautier C-H, Boudaoud A (2005) From Bouncing to Floating: Noncoalescence of Drops on a Fluid Bath. Physical Review Letters 94:177801

\bibitem[Dong et al.(2013)]{Dong2013} Dong X, Zhu H, Yang X (2013) Three-dimensional imaging system for analyses of dynamic droplet impaction and deposit formation on leaves. Transactions of the ASABE 56:1641-1651

\bibitem[Dong et al.(2014)]{Dong2014} Dong X, Zhu H, Yang X (2014) Characterization of droplet impact and deposit formation on leaf surfaces. Pest Management Science, doi: 10.1002/ps.3806

\bibitem[Dorr et al.(2008)]{Dorr2008} Dorr G, Hanan J, Adkins S, Hewitt A, O'Donnell C, Noller B (2008) Spray deposition on plant surfaces: a modelling approach. Functional Plant Biology 35:988-996

\bibitem[Dorr et al.(2013)]{Dorr2013} Dorr GJ, Hewitt AJ, Adkins S, Hanan J, Noller B (2013) A comparison of initial spray characteristics produced by agricultural nozzles. Crop Protection 53:109-117

\bibitem[Dorr et al.(2014)]{Dorr2013a} Dorr GJ, Kempthorne DM, Mayo LC, Forster WA, Zabkiewicz JA, McCue SW, Belward JA, Turner IW, Hanan J (2014) Towards a model of spray--canopy interactions: Interception, shatter, bounce and retention of droplets on horizontal leaves. Ecological Modelling 290:94-101.

\bibitem[Fitt et al.(1989)]{Fitt1989} Fitt BDL, McCartney HA, Walklate PJ (1989) The role of rain in dispersal of pathogen inoculum. Ann. Rev. Pytopathol. 27:241-270

\bibitem[Forster et al.(2005)]{Forster2005} Forster WA, Kimberley MO, Zabkiewicz JA (2005) A universal spray droplet adhesion model. Transactions of the ASAE 48:1321-1330

\bibitem[Forster et al.(2010)]{Forster2010} Forster WA, Mercer GN, Schou WC (2010) Process-driven models for spray droplet shatter, adhesion or bounce. 9th International Symposium on Adjuvants for Agrochemicals 277-285

\bibitem[Fox et al.(1992)]{Fox1992} Fox RD, Reichard DL, Brazee RD (1992) A video analysis system for measuring droplet motion. Applied Engineering in Agriculture 8:153-157

\bibitem[Gigot et al.(2014)]{Gigot2014} Gigot C, de Vallavieille-Pope C, Huber L, Saint-Jean S (2014) Annals of Botany 114:863–875

\bibitem[Gilet and Bush(2012)]{Gilet2012} Gilet T, Bush JW (2012) Droplets bouncing on a wet, inclined surface. Phys. Fluids 24:122103

\bibitem[Holloway(1970)]{Holloway1970} Holloway PJ (1970) Surface factors affecting the wetting of leaves. Pesticide Science 1:156-163

\bibitem[Hua and Rosen(1988)]{Hua1988} Hua XY, Rosen MJ (1988) Dynamic surface tension of aqueous surfactant solutions: I. Basic paremeters. Journal of Colloid and Interface Science 124:652-659

\bibitem[Huber et al.(1997)]{Huber1997} Huber L, McCartney HA, Fitt BDL (1997) Influence of target characteristics on the amount of water splashed by impacting drops. Agricultural and Forest Meteorology 87:201-211

\bibitem[Kang and Ng(2006)]{Kang2006} Kang CW, Ng HW (2006) Splat morphology and spreading behavior due to oblique impact of droplets onto substrates in plasma spray coating process. Surface and Coatings Technology 200:5462-5477

\bibitem[Kim and Chun(2001)]{Kim2001} Kim H-Y, Chun J-H (2001) The recoiling of liquid droplets upon collision with solid surfaces. Physics of Fluids 13:643-659

\bibitem[Kolinski et al.(2014)]{Kolinski2014} Kolinski JM, Mahadevan L, Rubinstein SM (2014) Lift-off instability during the impact of a drop on a solid surface. Physical Review Letters 112:134501

\bibitem[Kolinski et al.(2014a)]{Kolinski2014a} Kolinski JM, Mahadevan L, Rubenstein SM (2014a) Drops can bounce from perfectly hydrophilic surfaces. Europhysics Letters 108:24001

\bibitem[Lee and Kim(2004)]{Lee2004} Lee HJ, Kim H-Y (2004) Control of rebound with solid target motion. Physics of Fluids 16:3715-3719

\bibitem[Liu et al.(2014)]{Liu2014} Liu Y, Moevius L, Xu X, Qian T, Yeomans JM, Wang Z (2014) Pancake bouncing on superhydrophobic surfaces. Nature Physics 10:515-519

\bibitem[Liu et al.(2015)]{Liu2015} Liu Y, Tan P, Xu L (2015) Kelvin-Helmholtz instability in an ultrathin air film causes drop splashing on smooth surfaces.  Proceedings of the National Academy of Science 112:3280-3284

\bibitem[Mack(1936)]{Mack1936} Mack GL (1936) The determination of contact angles from measurements of the dimensions of small bubbles and drops. I. The spheroidal segment method for acute angles. J. Phys. Chem 40:159-167

\bibitem[Malgarinos et al.(2014)]{Malgarinos2014} Malgarinos I, Nikolopoulos N, Marengo M, Antonini C, Gavaises M (2014) VOF simulations of the contact angle dynamics during the drop spreading: Standard models and a new wetting force model. Colliod and Interface Science 212:1-20

\bibitem[Mandre and Brenner(2012)]{Mandre2012} Mandre S, Brenner MP (2012) The mechanism of a splash on a dry solid surface. Journal of Fluid Mechanics 690:148-172

\bibitem[Mangili et al.(2012)]{Mangili2012} Mangili S, Antonini C, Marengo M, Amirfazli A (2012) Understanding the drop impact phenomenon on soft PDMS substrates. Soft Matter 8:10045-10054

\bibitem[Mao et al.(1997)]{Mao1997} Mao T, Kuhn D, Tran H (1997) Spread and rebound of liquid droplets upon impact on flat surfaces. AIChE Journal 43:2169-2179

\bibitem[Marengo et al.(2011)]{Marengo2011} Marengo M, Antonini C, Roisman IV, Tropea C (2011) Drop collisions with simple and complex surfaces. Current Opinion in Colloid \& Interface Science 16:292-302

\bibitem[Massinon and Lebeau(2012)]{Massinon2012} Massinon M, Lebeau F (2012) Experimental method for the assessment of agricultural spray retention based on high-speed imaging of drop impact on a synthetic superhydrophobic surface. Biosystems Engineering 112:56-64

\bibitem[Massinon et al.(2014)]{Massinon2014} Massinon M, Boukhalfa H, Lebeau F (2014) The effect of surface orientation on spray retention. Precision Agriculture 15:241-254

\bibitem[Mayo et al.(2015)]{Mayo2015} Mayo LC, McCue SW, Moroney TJ, Forster WA, Kempthorne DM, Belward JA, Turner IW (2015) Simulating droplet motion on virtual leaf surfaces.  Royal Society Open Science 2:140528

\bibitem[Mercer et al.(2010)]{Mercer2010} Mercer GN, Sweatman WL, Forster WA (2010) A model for spray droplet adhesion, bounce or shatter at a crop leaf surface. Progress in Industrial Mathematics at ECMI 2008 945-951

\bibitem[Moevius et al.(2014)]{Moevius2014} Moevius L, Liu Y, Wang Z, Yeomans JM (2014) Pancake bouncing: Simulations and theory and experimental verification. Langmuir 30:13021-13032

\bibitem[Moreira and Moita(2010)]{Moreira2010} Moreira ALN, Moita AS, Panao MR (2010) Advances and challenges in explaining fuel spray impingement: How much of single droplet impact research is useful? Progress in Energy and Combustion Science 36:554-580

\bibitem[Mundo et al.(1995)]{Mundo1995} Mundo C, Sommerfeld M, Tropea C (1995) Droplet-wall collisions: Experimental studies of the deformation and breakup process. Int. J. Multiphase Flow 21:151-173

\bibitem[Nairn et al.(2011)]{Nairn2011} Nairn JJ, Forster WA, van Leeuwen RM (2011) Quantification of physical (roughness) and chemical (dielectric constant) leaf surface properties relevant to wettability and adhesion. Pest Management Science 67:1562-1570

\bibitem[Nairn et al.(2014)]{Nairn2014} Nairn JJ, Forster WA, van Leeuwen RM (2014) Influence of spray formulation surface tension on spray droplet adhesion and shatter on hairy leaves. New Zealand Plant Protection 67:278-283

\bibitem[Nairn et al.(2015)]{Nairn2015} Nairn JJ, Forster WA, Van Leeuwen RM (2015) Effect of solution and leaf surface polarity on droplet spread area and contact angle.  Accepted for publication, Pest Management Science.

\bibitem[Nikolov et al.(2002)]{Nikolov2002} Nikolov AD, Wasan DT, Chengara A, Koczo K, Policello GA, Kolossvary I (2002) Superspreading driven by Marangoni flow. Advances in Colloid and Interface Science 96:325-338

\bibitem[Okumura et al.(2003)]{Okumura2003} Okumura K, Chevy F, Richard D, Qu\'{e}r\'{e} D. Clanet C (2003) Water spring: A model for bouncing drops. Europhysics Letters 62:237-243

\bibitem[Pepper et al.(2008)]{Pepper2008} Pepper RE, Courbin L, Stone HA (2008) Splashing on elastic membranes: The importance of early time dynamics. Physics of Fluids 20:082103

\bibitem[Reichard et al.(1986)]{Reichard1986} Reichard DL, Brazee RD, Bukovac MJ, Fox RD (1986) A System for photographically studying droplet impaction on leaf surfaces. Transactions of the ASAE 29:707-713

\bibitem[Reichard et al.(1998)]{Reichard1998} Reichard DL, Cooper JA, Bukovac MJ, Fox RD (1998) Using a Videographic System to assess spray droplet impaction and reflection from leaf and artificial surfaces. Pesticide Science 53:291-299

\bibitem[Rein and Delplanque(2008)]{Rein2008} Rein M, Delplanque J-P (2008) The role of air entrainment on the outcome of drop impact on a solid surface. Acta Mechanica 201:105-118

\bibitem[Riboux and Gordillo(2014)]{Riboux2014} Riboux G, Gordillo JM (2014) Experiments of drops impacting a smooth solid surface: A model of the critical impact speed for drop splashing. Physical Review letters 113:024507

\bibitem[Richard and Qu\'{e}r\'{e}(2000)]{Richard2000} Richard D, Qu\'{e}r\'{e} D (2000) Bouncing water drops. Europhysics Letters 50:769-775

\bibitem[Rioboo et al.(2001)]{Rioboo2001} Rioboo R, Tropea C, Marengo M (2001) Outcomes from a drop impact on solid surfaces. Atomizations and Sprays 11:155-165

\bibitem[Roux and Cooper-White(2004)]{Roux2004} Roux DCD, Cooper-White JJ (2004) Dynamics of water spreading on a glass surface. Journal of Colloid and Interface Science 277:424–436

\bibitem[de Ruiter et a.(2015)]{Ruiter2015} de Ruiter J, Lagraauw R, van den Ende D, Mugele F (2015) Wettability-independent bouncing on flat surfaces mediated by thin air films.  Nature Physics 11:48–53

\bibitem[Saint-Jean et al.(2006)]{SaintJean2006} Saint-Jean S, Testa A, Madden LV, Huber L (2006) Relationship between pathogen splash dispersal gradient and Weber number of impacting drops. Agricultural and Forest Meteorology 141:257-262

\bibitem[Saint-Jean et al.(2004)]{SaintJean2004} Saint-Jean S, Chelle M, Huber L (2004) Agricultural and Forest Meteorology 121:183-196

\bibitem[Stalder et al.(2010)]{Stalder2010} Stalder AF, Melchior T, M\"{u}ller M, Sage D, Blu T, Unser M (2010) Low-Bond Axisymmetric Drop Shape Analysis for Surface Tension and Contact Angle Measurements of Sessile Drops. Colloids and Surfaces A: Physicochemical and Engineering Aspects 364:72-81

\bibitem[Stevens et al.(1993)]{Stevens1993} Stevens PJG, Kimberley MO, Murphy DS, Policello GA (1993) Adhesion of spray droplets to foliage: the role of dynamic surface tension and advantages of organosilicone surfactants.  Pesticide Science 38:237-245

\bibitem[Taylor(2011)]{Taylor2011} Taylor P (2011) The wetting of leaf surfaces. Current Opinion in Colloid and Interface Science 16:326-334

\bibitem[Vander Wal et al.(2006)]{VanderWal2006} Vander Wal RL, Berger GM, Mozes SD (2006) The splash/non-splash boundary upon a dry surface and thin fluid film. Experiments in Fluids 40:53-59

\bibitem[Venzmer(2011)]{Venzmer2011} Venzmer J (2011) Superspreading – 20 years of physicochemical research. Current Opinion in Colloid and Interface Science 16:335-343

\bibitem[Veremieiev et al.(2014)]{Veremieiev2014} Veremieiev S, Brown A, Gaskell PH, Glass CR, Kapur N, Thompson HM (2014) Modelling the flow of droplets of bio-pesticide on foliage.  Interfacial Phenomena and Heat Transfer 2:1-14

\bibitem[Wang et al.(2009)]{Wang2009} Wang MJ, Lin FH, Ong JY, Lin SY (2009) Dynamic behaviors of droplet impact and spreading -- Water on glass and paraffin. Colloids and Surfaces A: Physicochem. Eng. Aspects 339:224–231

\bibitem[Wang et al.(2013)]{Wang2013} Wang X, Chen L, Bonaccurso E, Venzmer J (2013) Dynamic wetting of hydrophobic polymers by aqueous surfactant and superspreader solutions. Langmuir 29:14855-14864

\bibitem[Wang et al.(2014)]{Wang2014} Wang S, He X, Song J, Zhang L, Dorr GJ, Herbst A (2014) Measurement comparison and fitted distribution equation of droplet size for agricultural nozzles. Transactions of the Chinese Society of Agricultural Engineering 30:34-42 (in Chinese with English abstract)

\bibitem[Wang et al.(2015)]{Wang2015} Wang S, Dorr GJ, Khashehchi M, He X (2015) Performance of selected agricultural spray nozzles using particle image velocimetry. Journal of Agricultural Sciences and Technology 17:601-613

\bibitem[Wirth et al.(1991)]{Wirth1991} Wirth W, Storp S, Jacobsen W (1991) Mechanisms controlling leaf retention of agricultural spray solutions. Pesticide Science 33:411-420

\bibitem[Xu et al.(2005)]{Xu2005} Xu L, Zhang WW, Nagel SR (2005) Drop splashing on a dry smooth surface. Physical Review Letters 94:184505

\bibitem[Xu et al.(2011)]{Xu2011} Xu LY, Zhu HP, Ozkan HE, Bagley WE, Krause CR (2011) Droplet evaporation and spread on waxy and hairy leaves associated with type and concentration of adjuvants. Pest Management Science 67:842-851

\bibitem[Yarin and Weiss(1995)]{Yarin1995} Yarin AL, Weiss DA (1995) Impact of drops on solid surfaces: self-similar capillary waves, and splashing as a new type of kinematic discontinuity. Journal of Fluid Mechanics 283:141-173

\bibitem[Yang(2012)]{Yang2012} Yang X (2012) Pesticide study on droplet deposition characteristics of typical crops. PhD Thesis, China Agricultural University, Beijing, China (in Chinese with English abstract)

\bibitem[Yarin(2006)]{Yarin2006} Yarin AL (2006) Drop impact dynamics: splashing, spreading, receding, bouncing. Annual Review of Fluid Mechanics 38:159-192

\bibitem[Yoon and Desjardin(2006)]{Yoon2006} Yoon SS, DesJardin PE (2006) Modelling spray impingement using linear stability theories for droplet shattering. International Journal for Numerical Methods in Fluids 50:469-489

\bibitem[Zen et al.(2010)]{Zen2010} Zen T, Chou F, Ma J (2010) Ethanol drop impact on an inclined moving surface. International Communications in Heat and Mass Transfer 37:1025-1030

\bibitem[Zhang et al.(2006)]{Zhang2006} Zhang Y, Zhang G, Han F (2006) The spreading and superspreading behaviour of new glucosamide-based trisiloxane surfactants on hydrophobic foliage. Colloids and Surfaces A: Physiochem. Eng. Aspects 276:100-106

\bibitem[Zwertvaegher et al.(2014)]{Zwertvaegher2014} Zwertvaegher IK, Verhaeghe M, Brusselman E, Verboven P, Lebeau F, Massinon M, Nicola\"{i} BM, David Nuyttens D (2014) The impact and retention of spray droplets on a horizontal hydrophobic surface. Biosystems Engineering 126:82-91

\end{thebibliography}
\end{document}